\documentclass{elsart}

\usepackage{natbib}
\usepackage{color}
\usepackage{graphicx}
\usepackage{epsfig}

\usepackage{amssymb}
\begin{document}

\begin{frontmatter}



\title{The Cosmic Foreground Explorer (COFE): A balloon-borne microwave polarimeter to characterize polarized foregrounds}


\author[ucsb,inpe]{Rodrigo Leonardi}, \author[ucsb]{Brian Williams}, \author[milano]{Marco Bersanelli}, 
\author[inpe]{Ivan Ferreira}, \author[ucsb]{Philip M. Lubin}, \author[ucsb]{Peter R. Meinhold}, 
\author[ucsb]{Hugh O'Neill}, \author[ucsb]{Nathan C.Stebor}, \author[iasf]{Fabrizio Villa}, 
\author[inpe]{Thyrso Villela}, \author[inpe]{Carlos A. Wuensche}

\address[ucsb]{Physics Department, University of California, Santa Barbara, CA 93106}
\address[inpe]{Instituto Nacional de Pesquisas Espaciais, Divis\~{a}o de Astrof\'{i}sica, Caixa Postal 515, 12227-010, S\~{a}o Jos\'{e} dos Campos, SP, Brazil}
\address[milano]{Dipartimento di Fisica, Universit\`{a} degli Studi di Milano, Via Celoria 16, 20133, Milan, Italy}
\address[iasf]{INAF - IASF Bologna, Via P. Gobetti, 101, 40129, Bologna, Italy}

\begin{abstract}
The COsmic Foreground Explorer (COFE) is a balloon-borne microwave
polarimeter designed to measure the low-frequency and low-$\ell$
characteristics of dominant diffuse polarized foregrounds. Short
duration balloon flights from the Northern and Southern
Hemispheres will allow the telescope to cover up to $80\%$ of the
sky with an expected sensitivity per pixel better than 100
$\mu\mathrm{K}/\mathrm{deg}^{2}$ from 10 GHz to 20 GHz. This is an
important effort toward characterizing the polarized foregrounds
for future CMB experiments, in particular the ones that aim to
detect primordial gravity wave signatures in the CMB polarization
angular power spectrum.
\end{abstract}

\begin{keyword}
cosmology: observations \sep cosmic microwave background \sep
polarization foregrounds
\end{keyword}

\end{frontmatter}
\section{Introduction}
\label{sec:introduction}

Measurement of polarization anisotropies in the Cosmic Microwave
Background (CMB) is one of the great challenges in cosmology
today. Very sensitive measurements of these anisotropies,
particularly at large angular scales, will provide unique
constraints on the influence of gravitational waves on the
production of structure in the very early Universe and information
on the epoch of reionization.

Several experiments are running or in the planning stages, and
long term development for a future space mission attacking CMB
polarization is underway. To date, nearly all of the effort has
been directed towards maximizing the number of detectors in the
focal plane to achieve the required sensitivity. Relatively little
work is going into sub-orbital efforts to constrain polarization
fluctuations at the largest angular scales, those most interesting
for their impact on understanding the inflationary epoch and
ionization history of the universe. This is primarily because of
an unproven perception that very low multipoles will not be
accessible to any but space-based missions. Indeed, large scale
polarization has been searched for with ground based experiments
over the last 30 years. The COsmic Foreground Explorer (COFE) is a
balloon-borne instrument to measure the low frequency and
low-$\ell$ characteristics of some dominant polarized foregrounds.
Good understanding of these foregrounds is critical both for
interpreting recent results, e.g. \cite{spergel06}, and for
appropriately planning future CMB missions. The experiment also
explores low-$\ell$ limits to CMB polarization measurements at
moderate frequencies from non-space based platforms. We believe
that balloon and ground-based measurements to characterize in
detail the polarized microwave sky are essential to prepare a
future space mission dedicated to CMB B-modes.

\section{Science}

The CMB radiation field is an observable that provides direct
information from the early Universe. The temperature and
polarization characteristics of this field impose constraints on
cosmological scenarios relevant to understand the origin and the
structure of the Universe. Accurate measurements of the CMB are
vital to improve our understanding about geometry, mass-energy
composition, and reionization of the Universe. Ultimately, the CMB
could also provide indirect detection of a stochastic
gravitational background and information from the inflationary
epoch itself. Having this big picture in mind, several CMB
experiments are now trying to constrain the tensor-to-scalar ratio
value and to detect the B-mode signature.

Among all practical limitations to primordial tensor amplitude
detection, contamination due diffuse microwave foreground
polarized emission is certainly the fundamental one. This emission
presents spatial and frequency variations that are not well known,
and the residuals from foreground subtraction are restricting our
knowledge of CMB polarization. This is particularly true for
future B-mode experiments that will benefit if accurate
determinations of spatial and spectral characteristics of
polarized foreground are made. For this reason, multifrequency
measurements of the polarized foregrounds in the microwaves is now
recognized as a key objective within the CMB community.

At low frequencies, foregrounds include synchrotron, free-free,
and possible spinning dust emission. Synchrotron dominates the low
frequency range of the microwave sky. Its emission is caused by
relativistic charged particles interacting with the Galactic
magnetic field and can be highly polarized. Synchrotron
measurements provide better understanding of the Galactic magnetic
field structure and the density of relativistic electrons across
the Galaxy. Free-free emission becomes more important in the
microwave intermediate frequency range, and it is due to
electron-ion scattering. Free-free is expected to be unpolarized
but this might not be true at the edges of HII clouds. Electrical
dipole emission from spinning dust has also been suggested by
recent observations at low microwave frequencies, e.g.
\cite{finkbeiner04}.

COFE is a balloon-borne microwave polarimeter to measure spatial
and low-frequency characteristics of diffuse polarized
foregrounds. This is an important effort toward characterizing the
polarized foregrounds for future CMB experiments, in particular
the ones that aim to detect primordial gravitational wave
signatures in the CMB polarization angular power spectrum.

\section{Instrumentation}

\subsection{Telescope}

\begin{figure}[!ht]
\begin{center}
\includegraphics[width=15.0cm]{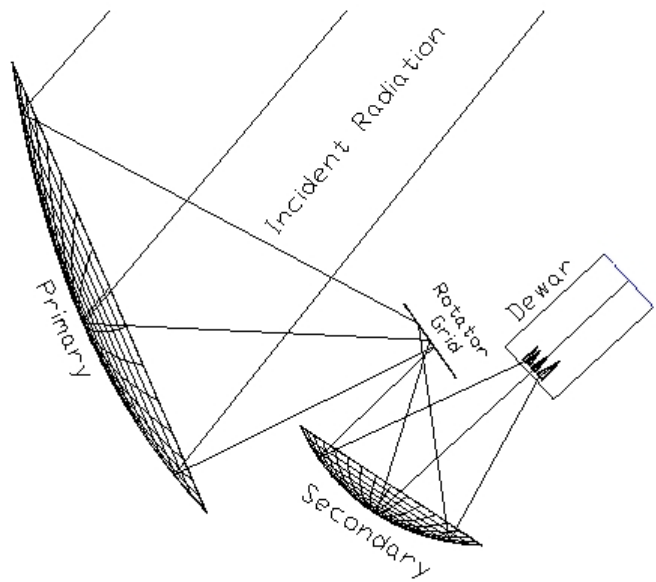}
\end{center}
\caption{Optical schematic for COFE and B-machine prototype
telescopes, an off-axis Gregorian configuration optimized for
minimal cross-polarization contamination. A $2.2$ m parabolic
reflector primary, a $0.9$ m ellipsoidal secondary, and a $0.3$ m
rotator grid are shown.} 
\label{fig:OpticalLayoutBW}
\end{figure}

A modified BEAST telescope design is the basis for the COFE optics
\citep{childers05, figueiredo05, meinhold05, mejia05, odwyer05}.
It consists of an off-axis Gregorian configuration obeying the
Dragone–Mizuguchi condition \citep{Drag78, Miz78}. The telescope
is optimized for minimal cross-polarization contamination and
maximum focal plane area. The primary reflector is a $2.2$ m
off-axis parabolic reflector. The incoming radiation is reflected
off of the primary reflector towards a polarization modulating
wave plate then to the secondary reflector. The $0.9$ m
ellipsoidal secondary reflects the incoming radiation toward the
array of scalar feed horns that couple the radiation to an array
of cryogenic low noise amplifiers. The telescope will be mounted
in a gondola that has been simplified from a standard
balloon-borne design due to the very light carbon fiber optical
elements. A schematic of the optics is shown in 
Figure \ref{fig:OpticalLayoutBW}.

\subsection{Polarization modulator}

\begin{figure}[!ht]
\begin{center}
\includegraphics[width=17.0cm]{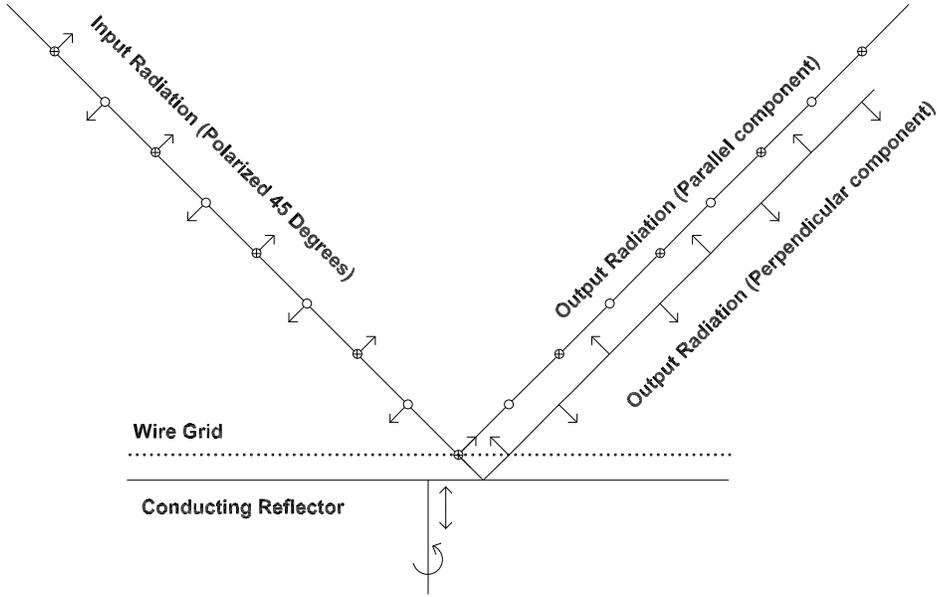}
\end{center}
\caption{Schematic of the polarization modulator. The input wave is
decomposed into its two linear polarization states, parallel and
perpendicular to the wires (represented by dots just above the
conducting reflector). The perpendicular component is phase shifted
from the extra path length. When added back to the parallel
component, the plane of polarization of the input wave is rotated.}
\label{fig:rotatorGrid}
\end{figure}

\begin{figure}[!ht]
\begin{center}
\includegraphics[width=10.0cm]{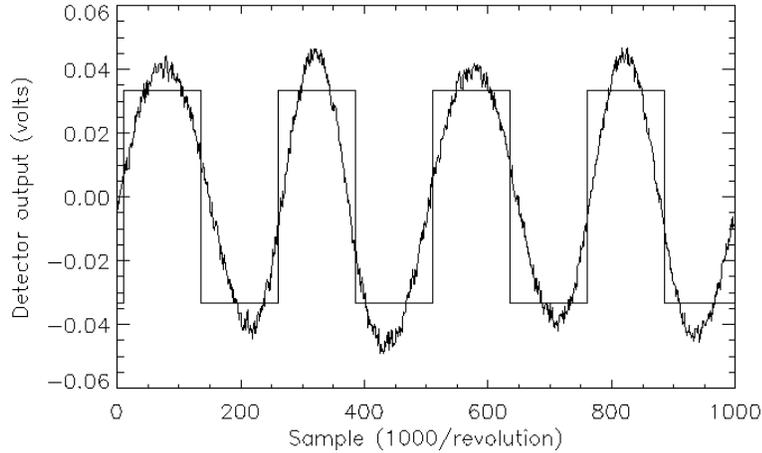}
\end{center}
\caption{Sample signal from a polarized thermal source. A single
revolution of the modulator is shown, along with the reference
signal to be used for demodulation. Commutating using this signal
yields $Q$, for instance, while demodulating with a reference phase
shifted by $\pi/4$ gives $U$.} 
\label{fig:signal}
\end{figure}

COFE will employ a low-loss reflective polarization modulator for
measuring both Q and U simultaneously. It consists of a linear
polarizing wire grid mounted in front of a reflecting plate. The
wire grid decomposes the input wave into components, parallel and
perpendicular to the wires, reflecting the parallel component with
low loss. The perpendicular component passes through the wire grid
and reflects off the back short, passes through the grid again and
recombines with the parallel component. The distance between the
plate and the grid introduces a phase shift between the two
components, effectively rotating the plane of polarization of the
input wave. A schematic of the polarization modulator is shown in
Figure \ref{fig:rotatorGrid}. 
Rotating the grid chops between the
two polarization states four times per revolution as shown in
Figure \ref{fig:signal}.

\begin{figure}[!ht]
\begin{center}
\includegraphics[width=12.0cm]{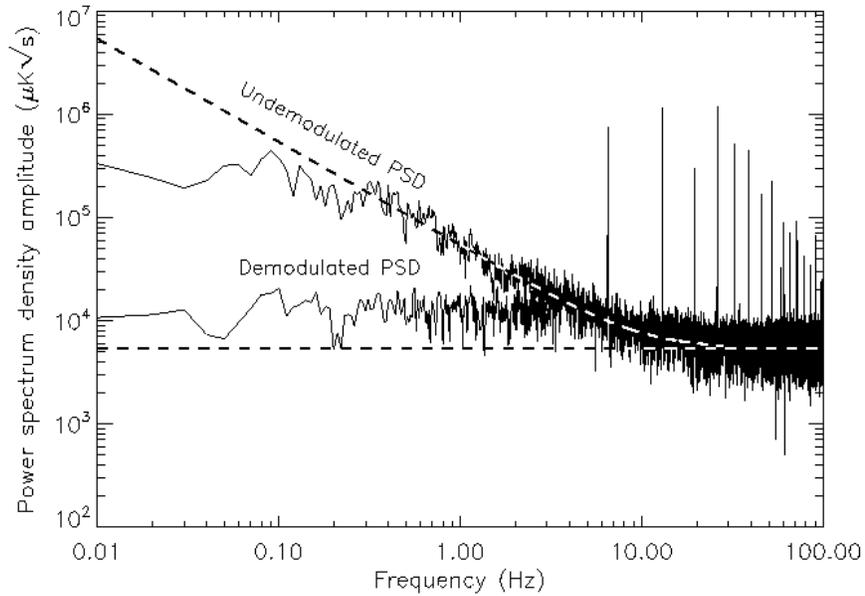}
\end{center}
\caption{Sample data from our room temperature radiometer viewing
the sky at 41.5 GHz. The undemodulated PSD displays the $1/f$ knee
of the HEMT radiometer of 10 Hz and a white noise of 5.4
$\mathrm{mK}\sqrt{\mathrm{s}}$. The demodulated data have no
visible $1/f$ and a white noise level consistent with
expectation.} 
\label{fig:ps}
\end{figure}

Tests of this modulator were performed at $41.5$ GHz, using a $70$
cm telescope. We measured beam patterns for the rotated
polarization states and integrated for extended periods on the sky
in Santa Barbara, CA. We were able to determine a $1/f$ knee lower
than $50$ mHz and very stable long term offsets. We also
demodulated sky data to the two different states and calculated
the correct combined sensitivity, as seen in Figure \ref{fig:ps}.

The polarization modulator has a broad bandwidth. We achieved 22
dB isolation at 20\% bandwidth. The radiometric loss of the
elements in the modulator can easily be made very low (of order
$0.1–1\%$) up to relatively high frequencies. The system works for
a very wide range of frequency bands.

\subsection{Receiver}

\begin{figure}[!ht]
\begin{center}
\includegraphics[width=15.0cm]{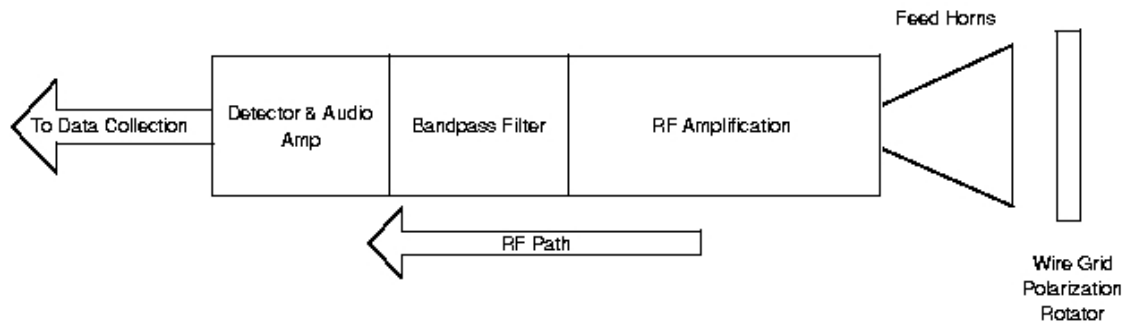}
\end{center}
\caption{Radiometer layout for COFE.}
\label{fig:Prototypedetector}
\end{figure}

COFE will use InP MMIC\footnote{Indium Phosphide Monolithic
Microwave Integrated Circuit} amplifiers integrated into simple
total power receivers. All of the RF gain will be integrated into
a small compact module inside the vacuum chamber. The module will
contain 3 to 4 amplifiers ($\sim75$ dB of gain), band pass filter,
cryogenic detector diode, and an audio amplifier. The module
avoids the need for cryo/vacuum waveguide feedthrus on the dewar
simplifying the overall design. The audio amplifiers will be
within the cryostat vacuum vessel for simplicity and noise
reasons, but will be at ambient temperatures. COFE has a modest
number of feeds required, and no orthomode transducers or hybrid
tees, so the passive components are minimal. A schematic of the
receiver is shown in Figure \ref{fig:Prototypedetector}.

\subsection{Data acquisition/demodulation}

Data acquisition will use the same technique we have been using in
our test system, namely synchronous sampling of analog
integrators. We oversample the data by a large factor and perform
the demodulation of $Q$ and $U$ Stokes parameters (and other modes
for systematic error analysis) in software. This yields the most
information and allows a variety of post-flight tests including
null signal analysis and analysis of the DC or total power
components (contaminated with $1/f$, but still useful for
systematic tests).

\subsection{Ground-based B-machine prototype}

A prototype polarimeter for a B-mode project, named B-machine, is
being deployed at the WMRS\footnote{White Mountain Research
Station} Barcroft facility, CA ($118^{\circ}14^{\prime}$ W
longitude, $37^{\circ}35^{\prime}$ N latitude, 3800 m altitude).
The WMRS facility is an excellent site for microwave observation
because of a cold microwave zenith temperature, low precipitable
water vapor, and a high percentage of clear days \citep{marvil06}.
Many of the components that will be used by the B-machine
prototype are useful for COFE as well. For example, the prototype
will allow systematic checks of the polarization modulator, and
COFE scan strategy. The B-machine prototype will be able to yield
some basic higher multipole results on the foregrounds as well as
the polarization signature and establish a data analysis pipeline.

\begin{figure}[!ht]
\begin{center}
\includegraphics[width=12.0cm]{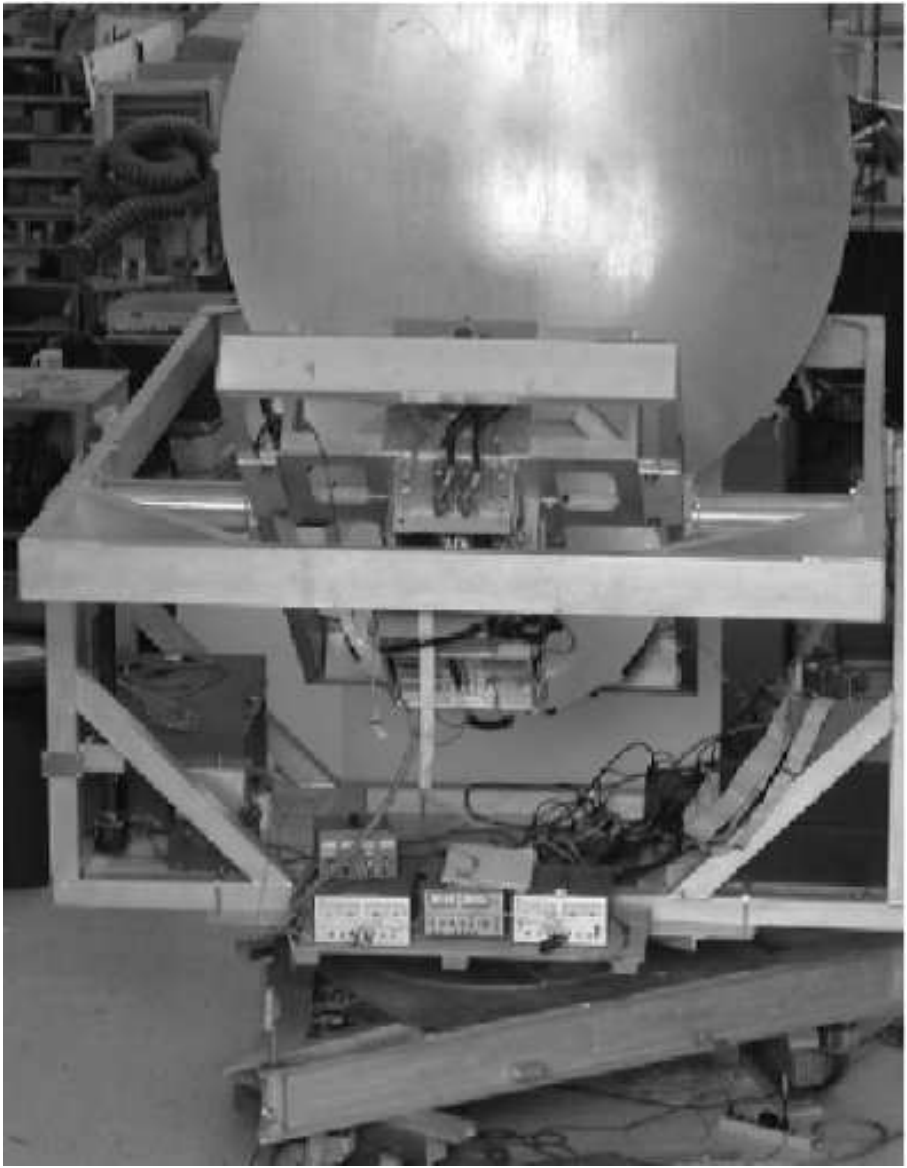}
\end{center}
\caption{Picture of prototype telescope to be deployed at WMRS.}
\label{fig:B-machinetelescope}
\end{figure}

The prototype possesses telescope and detector technology
identical to COFE. It has $2$ Ka-band and $6$ Q-band channels
centered at $31$ and $41.5$ GHz with FWHM resolution of
$28^\prime$ and $20^\prime$ respectively. The receiver has been
previously used in anisotropy measurements \citep{childers05}. The
telescope runs at constant elevation while continuously scanning
the sky in azimuth. A photograph of B-machine prototype is shown
in Figure \ref{fig:B-machinetelescope}.

\section{Performance}

\begin{figure}[!ht]
\begin{center}
\includegraphics[width=14.0cm]{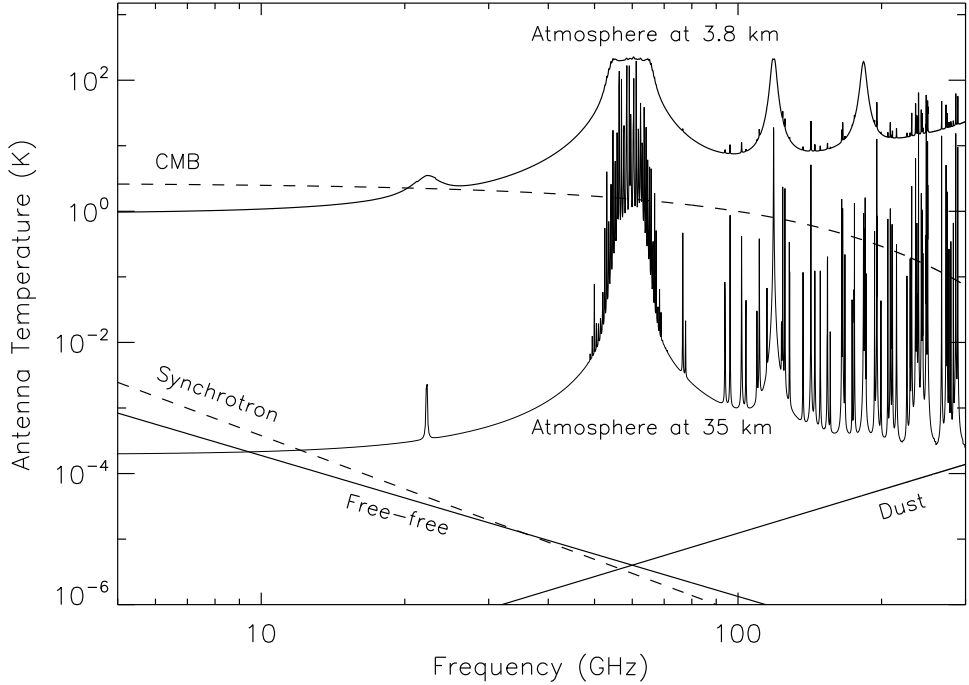}
\end{center}
\caption{Atmosphere, CMB, and predicted foreground emission from
$5$ to $300$ GHz. COFE bands run from $10$ to $20$ GHz. The zenith
atmosphere emission is shown at $3.8$ and $35$ km. The atmospheric
emission and lines are mainly due to $\mathrm{H}_{2}\mathrm{O}$,
$\mathrm{O}_{2}$, and $\mathrm{O}_{3}$. For the target altitude of
$35$ km, we expect well under $1$ mK total emission from the
atmosphere. Foreground spectral index $\beta$ for free-free,
synchrotron, and dust were assumed, respectively, as $-2.15$,
$-2.7$, and $2.2$.} 
\label{fig:atmosphere}
\end{figure}

For any sub-orbital CMB experiment, minimizing atmospheric
contamination is important. For the COFE bands, total atmospheric
emission at our target altitude of $35$ km is less than $1$ mK.
Common broad band bolometric atmospheric antenna temperature
contributions at balloon altitudes are several hundred mK or more.
Since the effective CMB antenna temperature drops with frequency,
our effective atmospheric signal is approximately $1000$ times
less than for a bolometric balloon-borne system. Hence low-$\ell$
information from a balloon-borne system is very clean by
comparison. Figure \ref{fig:atmosphere} 
shows the atmosphere and
predicted foreground emission over a range of frequencies
interesting for CMB work (the foreground prediction is calculated
from \cite{bennett03}).

\subsection{Receiver bands and expected receiver sensitivity}

Receiver sensitivity can be estimated according to the radiometer
equation
\begin{equation}\label{e:sensitivity}
   \sigma_{T}=K\left(\frac{T_{\mathrm{sys}}+T_{\mathrm{sky}}}{\sqrt{\Delta\nu\cdot\tau}}\right),
\end{equation}
where $ \sigma_{T}$ is the root-mean-square noise,
$T_{\mathrm{sys}}$ is the system noise temperature,
$T_{\mathrm{sky}}$ is the sky antenna temperature, $\Delta\nu$ is
the bandwidth, $\tau$ is the integration time, and $K$ is the
sensitivity constant of the receiver.

For COFE and B-machine prototype the sensitivity constant of each
receiver is $K=\frac{\pi}{2}$. The signal is sine wave modulated,
reducing the sensitivity by a factor of $\frac{\pi}{2}$ as compared
with a standard Dicke receiver, with an addition factor of
$\frac{1}{2}$ from the the standard definition for $Q$ and $U$ in
the Rayleigh-Jeans regime of the CMB spectrum. Table 1 shows our
estimation of the sensitivity of each receiver.

\begin{center}
Table 1 -- Instrument parameters.
\begin{tabular}{|l||r|r|r||r|r|r|}
  \hline
  &\multicolumn{3}{c||}{COFE}&\multicolumn{2}{c|}{B-machine}\\
  \hline
  \hline
  Central frequency (GHz) & 10 & 15 & 20& 31 & 41.5\\
  \hline
  FWHM beam (arcmin)  & 83 & 55 & 42 & 28 & 20\\
  \hline
  $T_{\mathrm{sys}}$ (K) & 8 & 10 & 12& 25 & 27\\
  \hline
  $T_{\mathrm{sky}}$ (K) at target altitude\footnotemark& 2.5 & 2.4 & 2.3& 6.4 & 13.0\\
  \hline
 Bandwidth (GHz) & 4 & 4 & 5& 10 & 7\\
  \hline
 Number of receivers & 3 & 6 & 10& 2 & 6\\
  \hline
 Sensitivity per receiver ($\mu \mathrm{K}\sqrt{\mathrm{s}})$
 & 261 & 308 & 318 & 493 & 751\\
  \hline
 Aggregate sensitivity $(\mu \mathrm{K}\sqrt{\mathrm{s}})$& 151 & 126 &
 100 & 348 & 307\\
 \hline
\end{tabular}
\end{center}
\footnotetext{For COFE and B-machine (ground based), we compute
expected $T_{\mathrm{sky}}$ antenna temperature at target altitude
of $35$ km and $3.8$ km, respectively.}

By increasing the number of receivers, future ground-based or
balloon-borne experiments can significantly improve aggregate
sensitivity. For instance, 30 detectors could reach 61 $\mu
\mathrm{K}\sqrt{\mathrm{s}}$ and 107 $\mu
\mathrm{K}\sqrt{\mathrm{s}}$ at 30 and 40 GHz, respectively.

\subsection{Scan strategy, sky coverage and expected map
sensitivity}

\begin{figure}[!ht]
\begin{center}
\includegraphics[width=12.0cm]{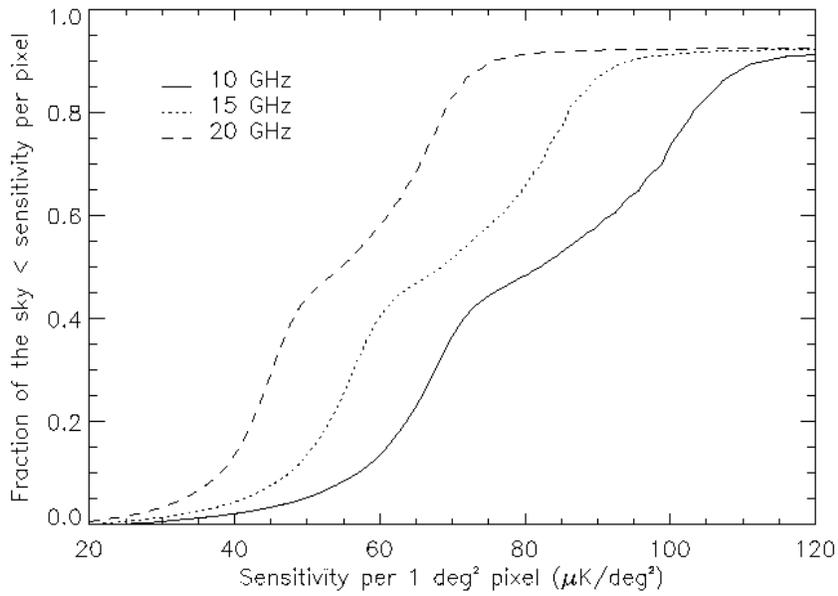}
\end{center}
\caption{Integrated histogram of anticipated aggregate sensitivity
per 1 $\mathrm{deg}^{2}$ pixel assuming a 24 hour flight from the
Northern Hemisphere (Fort Sumner, NM) and a 24 hour flight from
the Southern Hemisphere (Alice Springs, Australia). For each COFE
band, we plot the fraction of the entire sky measured with better
than a given aggregate sensitivity. The change of the curves slope
is due to the fact that $35\%$ of the sky can be observed from
both hemispheres using COFE scan strategy.} 
\label{fig:SPP}
\end{figure}

\begin{figure}[!ht]
\begin{center}
\includegraphics[angle=90,width=15.0cm]{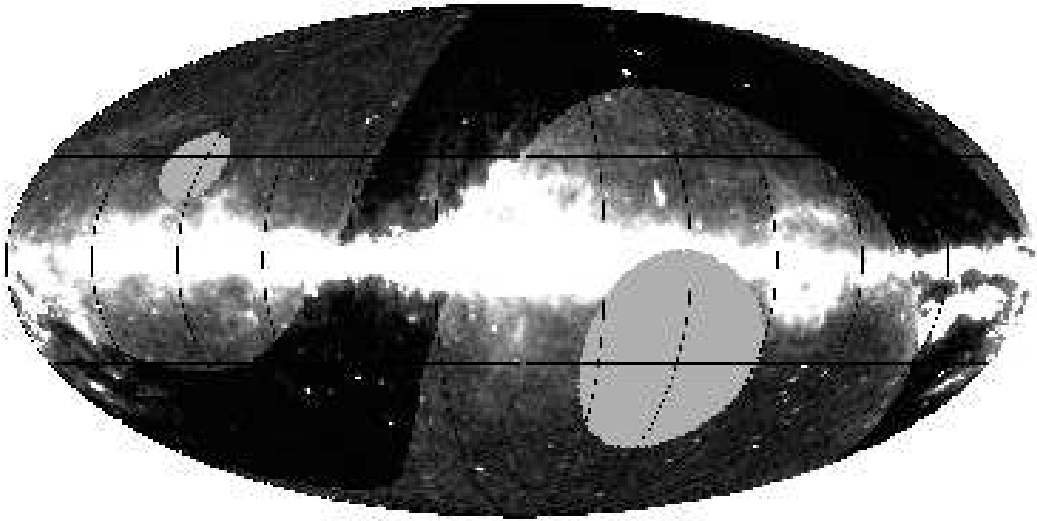}
\end{center}
\caption{Sky coverage for COFE assuming a 24 hour flight from the
Northern Hemisphere (Fort Sumner, NM) and a 24 hour flight from
the Southern Hemisphere (Alice Springs, Australia). The region
observed contains nearly the entire sky ($93\%$). The darker strip
shows the overlap between the two observations. For illustration
purposes, we show the diffuse Galactic structure obtained adding
synchrotron, free-free and dust maps at 23 GHz \citep{bennett03}.}
\label{fig:sky_coverage}
\end{figure}

COFE uses a simple scan strategy to cover the largest available
sky area in each flight. The telescope will be pointed nominally
45$^{\circ}$ from the horizon to minimize ground and balloon
pickup, and the gondola will rotate constantly at approximately
$1/2$ rpm. Data acquisition sample rate will be synchronized with
the polarization rotator (at $\sim 30$ Hz). For instance, using
this strategy, a $24$ hour flight from Fort Sumner, NM, allows to
cover $59\%$ of the sky area with a median aggregate pixel
sensitivity of 92 $\mu\mathrm{K}/\mathrm{deg}^{2}$, 77
$\mu\mathrm{K}/\mathrm{deg}^{2}$, and 61
$\mu\mathrm{K}/\mathrm{deg}^{2}$ at 10 GHz, 15 GHz, and 20 GHz
respectively. COFE will acquire data from nearly all of the sky
($\sim93\%$). This will be achieved in a set of $12$ and/or $24$
hour flights from the Northern and Southern Hemispheres. Figure \ref{fig:SPP} 
provides estimates for sensitivity per square degree
pixel over the whole sky for our flight plans. Figure \ref{fig:sky_coverage} 
illustrates the expected sky coverage.

The B-machine prototype focuses on higher multipoles but uses a
similar scanning strategy from the ground. For a conservative 60
day observing campaign at WMRS, we expect to cover $56\%$ of the
sky with an median aggregate sensitivity of 27
$\mu\mathrm{K}/\mathrm{deg}^{2}$, and 23
$\mu\mathrm{K}/\mathrm{deg}^{2}$ at 31 GHz and 41.5 GHz,
respectively.

\section{Conclusion}

Over the next few years we will field a balloon-borne telescope to
map more than $90\%$ of the sky. Both polarization anisotropy and
polarized foregrounds will be measured over several bands. This is
an important effort toward characterizing the polarized
foregrounds for future CMB experiments.

In addition to foreground detection, COFE will better characterize
the polarization modulation capability for measuring $Q$ and $U$
simultaneously. As discussed earlier, a large scale ground-based
campaign will capitalize on the technology that has been developed
by COFE and B-machine prototype.

It is clear that our current understanding of the polarization
foregrounds limits our ability to make accurate observations of
the B-mode signature. COFE will lessen the effect that incomplete
models of foregrounds will have on future experiments.

\section{Acknowledgments}

We acknowledge support from the National Aeronautics and Space
Administration (NASA), and the California Space Institute
(CalSpace). T.V. and C.A.W. acknowledge CNPq Grants 305219/2004-9
and 307433/2004-8, respectively. Some of the results have been
derived using the HEALPix\footnote{http://healpix.jpl.nasa.gov}
\citep{gorski05} package.




\end{document}